# Unveiling the mystery of nucleation and growth of carbon nanotube and layered graphene inside carbon arc-discharge

**Soumen Karmakar**[1,*]

[1]Department of Physics, Birla Institute of Technology, Mesra, Off-Campus Deoghar, Deoghar 814142, Jharkhand, INDIA
[*]Corresponding author.
E-mail address: skarmakar@bitmesra.ac.in

**A model for the formation of carbon nanotubes (CNTs) and layered graphene in an arc discharge method is developed on the basis of observed erosion of graphite anode under various experimental conditions and analyses of the morphology of the eroded anode-surface, concerned cathode deposits and their constituents. It is predicted that, cold thermal shock, triggered by the rapid movement of the anode-spot, leads to crack microbranching at some selected locations on the anode-surface. These crack-microbranches cleave and fragment the basal planes of pairs of adjacent crystallites into curved graphitic nanoribbons with minimum two basal planes. These nanoribbons further react chemically with the $C_2$ and $C_3$ precursors present in the dusty carbon plasma and evolve either in the form of CNT or layered-graphene, depending on the viscosity and composition of the gaseous environment they are exposed to, just after getting detached from the anode.**

## Introduction

CNTs were first discovered inside the soft core of the deposit formed on the cathode surface in a graphite electric-arc process [1]. Since then, they have widely been experimented in a variety of fields due to their outstanding electrical, thermal, chemical and mechanical properties [2]. Similar carbon arcs have also been known to produce few-layer graphene sheets (FLGs) as well [3]. These sheets are also equally potent, like CNTs, in many technological applications [4]. A number of models have evolved to understand the growth mechanism of both CNT and FLG in an arc-plasma process [3, 5-10]. However, carbon electric-arcs still remain highly mysterious, as it is not yet well understood how CNTs and FLGs nucleate and grow in such a process. As a result, in the arc-discharge process, CNTs and FLGs are known to be rather by-products, as the major fraction (as high as >90%) of the consumed carbon source (conventionally, commercial graphite) is converted into amorphous carbon, nanoparticles, graphitic chunks and flaky structures. Because of lack of proper knowledge about the growth mechanism, it is therefore not yet possible to optimize the arc process to its best, for the tailored synthesis of CNTs and FLGs, and use these potent pristine structures for commercial applications, especially in composite science and technology.

Considering the high temperature (~ 4000 °C) of a carbon-arc, suitable for synthesizing CNTs, it is natural for carbon to take its most thermodynamically-stable form [11]. However, to the best of author's knowledge, there is no theoretical clue available in the literature as to this most stable form. The dependency of the internal structure of the anode, on the formation of CNTs [12], not only contravenes the theories based on gas-phase condensation [5], models [8, 9], based on liquid-phase-growth also remains highly controversial; since liquid phase of graphite has not yet been realized near atmospheric pressure even with sophisticated experimentations [13, 14].

In addition, no existing theory or model provides any concrete insight that can satisfactorily explain any of the following issues: (i) reason behind the incapability of a carbon arc to produce either single-walled CNTs or graphene sheets without using catalyst(s), (ii) why the inner diameter of such CNTs are found decreasing with the increase in number of walls, (iii) reason behind the observed variation in the diameter of such CNTs, (iv) why the generation of such CNTs are affected by the crystallographic structure of the anode, (v) the most fundamental property of the arc that decides the formation of both CNT and FLG, etc.

The present article presents a concept, which to the best of the knowledge of the author, is hitherto unknown to the scientific community. The model holds the potential to unambiguously explain most of the experimentally observed phenomena related to the electric-arc synthesized CNT and FLG.

## Materials and methods

A constant voltage DC electric-arc, in between two mirror polished, co-axial, cylindrical, spectroscopic 99.999% pure polycrystalline graphite (Graphite (India) Ltd.) electrodes with density 1.8 gcm$^{-3}$, housed inside a reactor, described elsewhere [3], was employed for the generation of data after removing the magnet-assembly. The diameters of the anode and cathode were 10 mm and 30 mm respectively. The water cooled grounded cathode was 50 mm long and the initial length of the anode was 70 mm. Both the electrodes were tightly fitted to two water-cooled copper holders, through which electrical connections were made. The buffer gas was 99.99% pure Ar, maintained at 500 ± 20 Torr during steady state of arc operation. Before each run, the reactor was evacuated by an oil rotary vacuum pump (ED 30, HHV, India) to a base pressure of 10$^{-3}$ Torr and then Ar was purged into the reactor. The arc was ignited by a 6 kW DC power supply (Ion Arc Machines (India) Pvt. Ltd) with open circuit voltage of 60 V by the 'touch and pull' method and the arc-voltage was kept constant at 20 ± 2 V. In order to keep the arc-voltage constant, the anode was constantly translated by a rack and pinion arrangement. The arc-current $I_{arc}$ was varied from 50 A to 160 A. The typical runtime was about 3 min for each run. The double walled stainless steel reactor chamber and the electrode-holders were constantly cooled by running water at the ambient temperature of 18 ˚C at a flow rate of 5 lpm. A thermocouple, with digital output, was placed about 10 cm away from the arc-zone, to record the temperature of the buffer gas ($T_b$). A couple of readings by the thermocouple were taken during each run and then the data were averaged to find out the most probable value of $T_b$.

After the completion of each operation, sufficient time was given to cool down the chamber to the room temperature and then the as synthesized cathode deposits (CDs) were dismantled from the top of the cathode, crushed thoroughly by a granite pestle and mortar and then investigated in their totality by Raman spectroscopy and transmission electron microscopy (TEM).

Raman spectroscopic analysis was carried out by a Jobin Yvon Labram HR800 spectrometer in the back-scattering geometry at room temperature, where a He-Ne laser with 1 mW power and 1 μm spot size was used. The data were recorded for the Raman shift within the range of 1000 cm$^{-1}$ to 2000 cm$^{-1}$. For each sample, three different measurements were carried out selecting three random portions of the sample. The recorded spectra were then base-line corrected and fitted with three Lorentzian line-shapes. The ratios of the areas under the bands appearing near 1580 cm$^{-1}$ (G-band) and 1320 cm$^{-1}$ (D-band) were estimated and then averaged for each sample.

TEM micrographs were captured using a 200 KV Tecnai G$^2$ 20 microscope. For this investigation, the powdered CDs (~ 50 mg) were mild sonicated in toluene (~ 10 ml) in separate ultra-clean glass bottles for about 5 min. The liquid samples, thus prepared, were then drop casted on carbon coated holey copper micro-grids and dried for a minute under an IR lamp. TEM images were recorded in the bright field geometry. 100 keV electron beam was employed for obtaining the electron diffraction patterns.

## Results

To understand the processes involved during arcing, first of all, some of the data generated by the electric-arc reactor, are analyzed. Fig. 1 demonstrates the variation in the reactor gas temperature ($T_g$) with arc-current ($I_{arc}$).

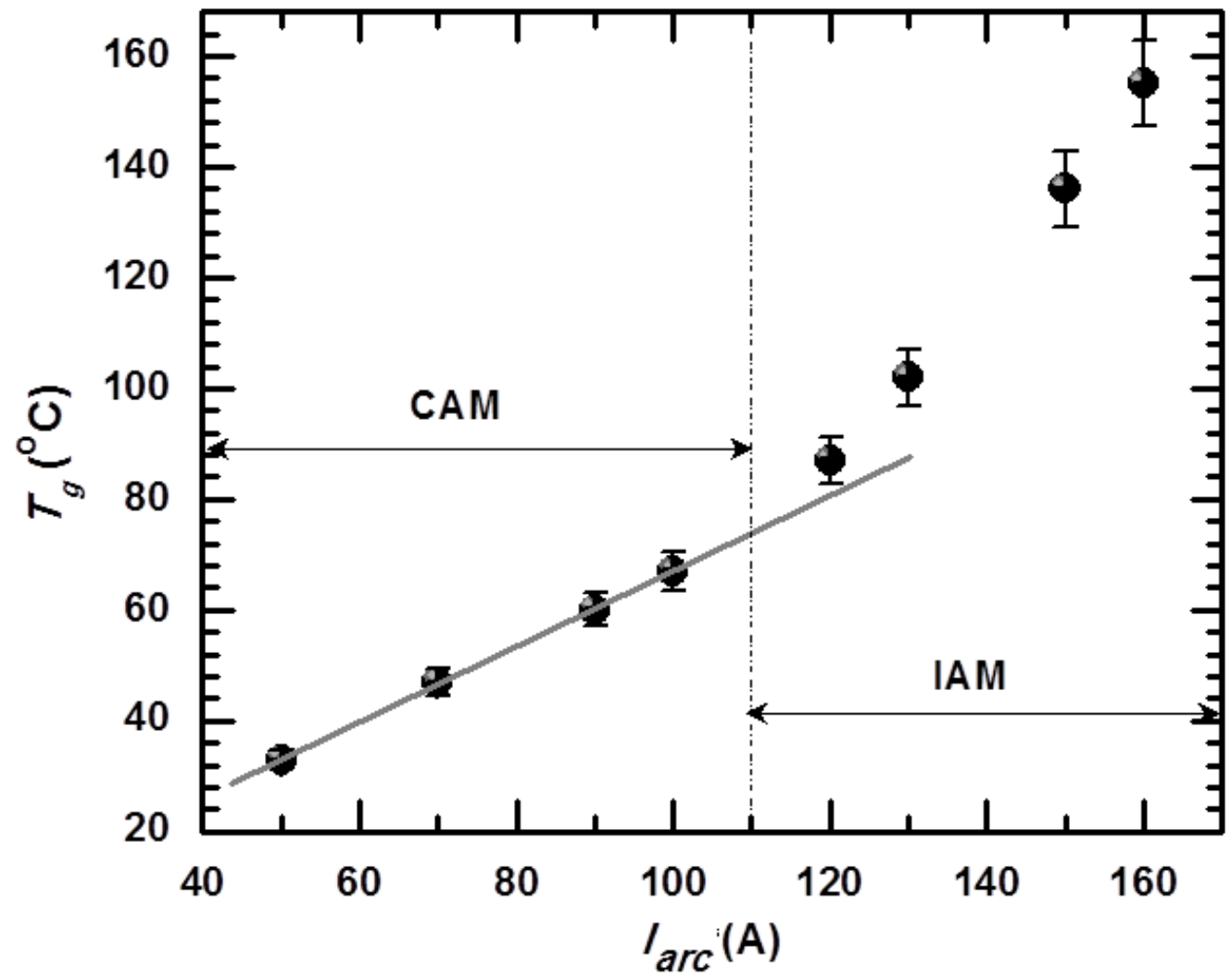

**Fig. 1** Variation of $T_g$ with $I_{arc}$. The linear relationship in between $T_g$ and $I_{arc}$ in the current range of 50 A – 100 A is highlighted by the solid straight line. The linear zone is attributed to CAM, while the non-linear portion corresponds to IAM.

It is interesting to note that the variation deviates from linearity as $I_{arc}$ is increased beyond ~ 100 A. It is quite obvious that this variation would closely mimic the variation in the Joule heat loss ($\dot{Q}_J$) from the arc column, per unit time. Hence, following conclusions may be drawn from Fig. 1.

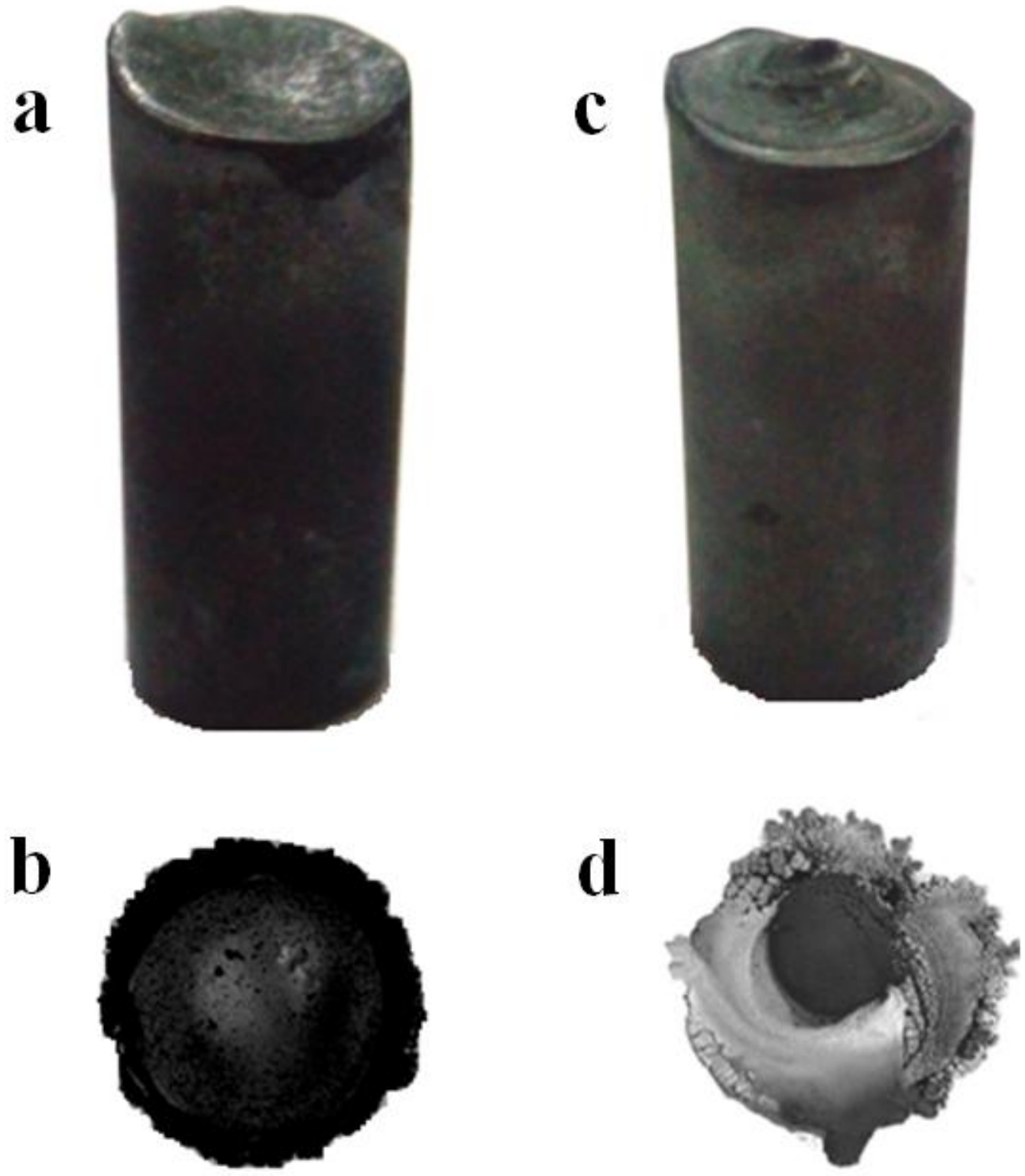

**Fig. 2** Typical photographs of the eroded anodes and CDs. Typical morphology of anode-surface after 3 min erosion under (a) CAM and (b) IAM. Typical top views of the CDs corresponding to CAM and IAM are illustrated in (c) and (d) respectively.

When $I_{arc}$ is increased from 50 A to 100 A, the arc is constricted and the cross-sectional areas of both the arc-channel and plasma column remain lesser than that of the anode and increase linearly with $I_{arc}$. In this current range, the current density $J_c$ of the arc-channel and the degree of ionization in the discharge do not vary to a fairly good approximation. Under such a condition, $\frac{T_g}{\dot{Q}_J}$ ratio is a constant. Let this mode be distinguished as constricted arc mode (CAM).

On the other hand, upon increasing $I_{arc}$ to 120 A, the arc transforms into an intense arc; in which, the plasma expands over the entire front surface of the anode with $T_p \geq T_s$ close to the anode. Here, $T_p$ and $T_s$ are the plasma temperature and graphite sublimation point respectively. Thus, the cross-sectional area of the anode is one of the parameters, which decides the threshold value of $I_{arc}$ at which CAM transforms into an intense arc mode (IAM). Once the arc enters into IAM, degree of ionization in the discharge tries to increase with $I_{arc}$ because of enhanced Joule heating and $T_g$ varies as $\frac{I_{arc}^2}{\tilde{\sigma}_{arc}}$, where $\tilde{\sigma}_{arc}$ is the volume averaged electrical conductivity of the arc-discharge. This non-linear variation is clearly visible in Fig.1.

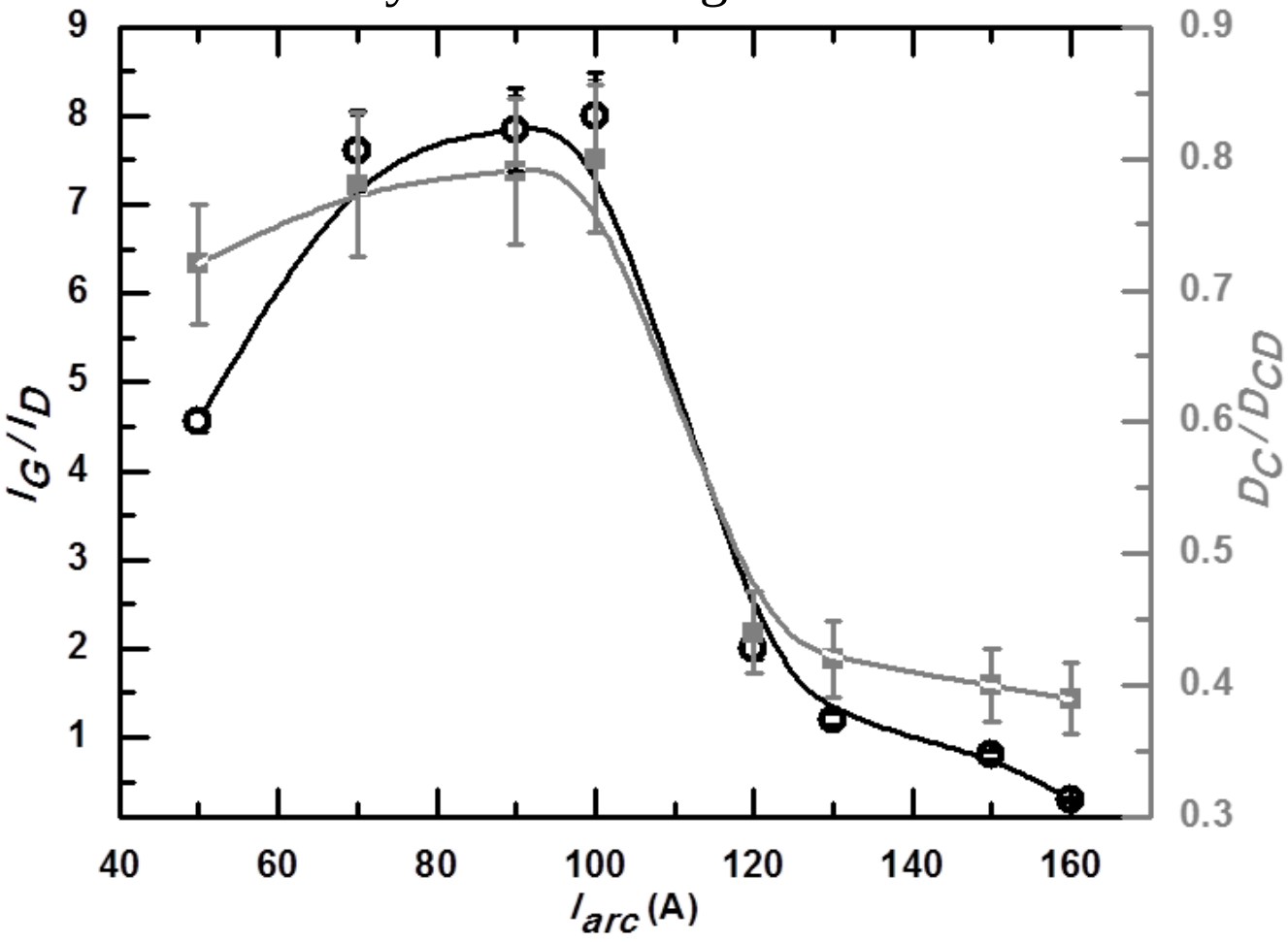

**Fig. 3** Variation in degree of graphitization of the CDs. The black circles represent the $\frac{I_G}{I_D}$ ratios, estimated from the first order Raman spectra; while the gray squares represent the $\frac{D_{CD}}{D_C}$ ratios. The solid lines are the B-Spine fits to the data to show their tentative variation with $I_{arc}$. The error bars represent the standard deviations of the data.

The transformation from CAM to IAM upon increasing $I_{arc}$ and the results of such a changeover may be analyzed from the surface morphologies of the eroded anodes and the structures of the associated CDs, as illustrated in Fig. 2. It is seen that, the axial region of the anode is eroded more than the peripheral zone; the anode-surface is smooth and concave after a CAM operation (Fig. 2a). The corresponding CDs (Fig. 2b) are uniform with a larger proportion of the core with respect to the outer shell. The situation reverses totally on running the arc under IAM. The anode is eroded more at the peripheral zone (Fig. 2c); its surface exhibits clear circular patch marks and the quantity of the hard outer shell is vividly much more than the core in the corresponding CDs (Fig. 2d).

The relative amounts of ordered graphene structured species in the cores of the CDs, synthesized at different values of $I_{arc}$ is presented in Fig. 3 in terms of the intensity ratio $\left(\frac{I_G}{I_D}\right)$ of the G to D bands [15], calculated from the respective Raman spectra. The variation exhibits a maxima near $I_{arc} = 100\,A$, beyond which a sharp fall in the ratio is noticed. The trend predicts that the amount of ordered structures comprising of graphene sheets is more inside the CDs generated by CAM than those produced by IAM.

The variation in the ratio of the diameter of the core ($D_c$) to that of the corresponding CD ($D_{CD}$), as a function of $I_{arc}$, is also shown in the same figure. Recalling that CNTs are formed only inside the cores, Fig. 3 predicts that CAM might assist in improving the conversion efficiency of the eroded mass into CNTs.

The morphologies of the ordered graphene structured species, as stated above, can be seen in the corresponding TEM micrographs. Fig. 4a is a typical TEM micrograph representing the samples synthesized under CAM operation, while Fig. 4b is the representative of the samples corresponding to IAM. In both the TEM images CNTs, carbon nano-crystalline particles (CNPs) and amorphous structures are common. However, Fig. 4b is characteristically different from Fig. 4a because of large amount of flaky structures present in it. While the electron diffraction pattern in Fig. 4a confirms the presence of CNTs within the samples synthesized under CAM, the diffraction pattern in Fig. 4b confirms the flakes to be turbostatic, planer graphitic structures [16]. While the samples generated under CAM are seen to be rich in CNTs, the flakes are plentiful in the samples produced by intense-arc.

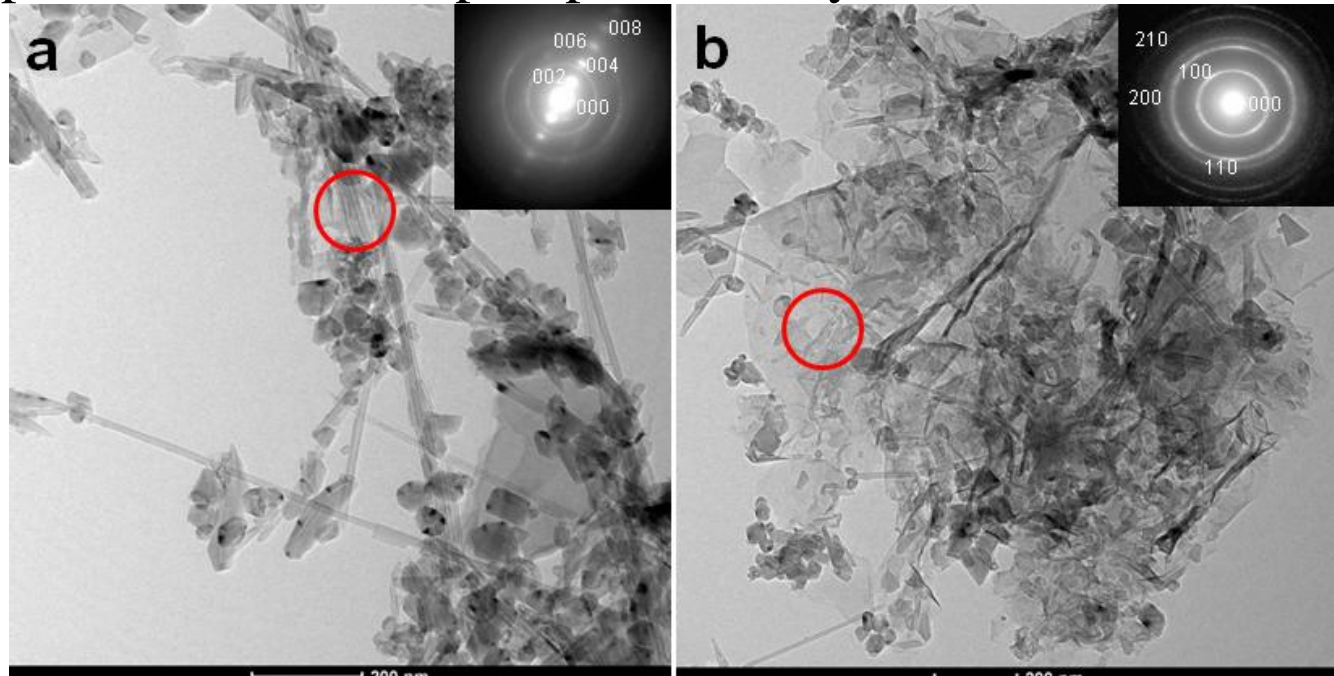

**Fig. 4** Typical bright-field TEM micrographs. The constituents of the CDs synthesized at $I_{arc}$= 100 A is shown in (a) and (b) represents the typical micrograph corresponding to $I_{arc}$= 160 A. The insets are the electron diffraction patterns, captured from the areas marked with the circles in (a) and (b). The diffraction patterns have been linked with their respective Miller indices, corresponding to pristine graphite.

## Discussion

Combining all the above findings, there remains little doubt that a carbon arc, operating in CAM is ideal for generating CNTs, while IAM gives rise mostly to flakes. To have a satisfactory answer to what presumably happens inside a carbon-arc, during the CAM and IAM operations, the effects of the input electrical power, electrodes-separation, nature of the buffer medium, structural deformation of the anode during arcing and the consequences thereafter require rigorous analysis.

Customarily, roughly oriented graphite, having frozen-in defective carbon layers along the grain boundaries, is used as anode for the synthesis of CNTs. Individual graphene in each crystallite is composed of numerous grains separated by quilt boundaries, where defects, e.g. voids and vacancies tend to segregate [17].

Suppose that, at steady state CAM, an arc at a constant chamber pressure $P_o$ runs in between two graphite electrodes maintained at a constant voltage difference ($V_{arc}$). For a short arc-length, which is ideal for the generation of CNTs [1], $V_{arc} \approx V_a + V_c$; where $V_a$ and $V_c$ are the anode and cathode voltage drops respectively. Under such a condition, gas ionization in the plasma column is negligible and $I_{arc}$ is mainly due to the flow of electrons. The huge amount of soot, deposited on the reactor chamber walls during arc-evaporation of graphite [15, 16], confirms the dominance of neutrals and diffusion process. Let the energy flux carried by the electrons in the arc channel to the anode be sufficient to initiate sublimation overriding the various energy losses. Under such a condition, the anode sublimates and a carbon a vapour-jet originates from the anode.

The vapour pressure, in front of the sublimating anode can best be described by the Clausius-Clapeyron equation:

$$ln\left(\frac{P_v}{P_p}\right) = \frac{\Delta H}{R}\left(\frac{1}{T_a} - \frac{1}{T_p}\right) \quad (1),$$

where, $P_v$ is the carbon vapour pressure in the vicinity of the eroding regions, $\Delta H$ is the heat of sublimation for graphite, $R$ is the universal gas constant and $T_p$ is measured adjacent to the anode-surface-region with a temperature $T_a$.

Near $T_s$, polycrystalline graphite behaves like a perfect thermal insulator [18]. Hence, during arcing under CAM, $T_a = T_s$ near the anode-spot and there are regions on the anode-surface with temperatures lower than $T_s$. The plasma column dissipates heat to the surrounding buffer gas with a much lower temperature and $T_{arc}$ and hence $\sigma_{arc}$ are the lowest at the plasma periphery. As a result, $I_{arc}$ flows mostly along the plasma centreline, where the values of $T_{arc}$ and $\sigma_{arc}$ are the maximum. As a result, the central part of the anode is eroded more than its peripheral zone. Fig. 2a confirms such an erosion process.

In contrary to CAM, while the anode sublimates under IAM, plasma facing surface of the anode attains uniform temperature $T_s$ and continues sublimating. Under such a condition, the arc becomes curved and the anode-spot anchors at the side surface of the anode [19]. With this arc configuration, the axial zone of the anode continues receiving heat from the plasma; while, the peripheral zone receives an excess energy flux from the impinging electrons. As a result, the peripheral region of the anode is eroded more than its axial zone, as is evident from Fig. 2b.

Under IAM configuration, following Eqn. 1, $T_p$ and hence, $P_p$ would be maximum at the plasma periphery; while the minimum values would correspond to the plasma centreline. Such a pressure gradient would give rise to the vortices, which would enhance the inflow of cooler buffer medium. As a result, plasma particles undergo rapid cooling and condense in the form of amorphous structures near the plasma boundary. With the increase in $I_{arc}$, the spatial extent of such rapid quenching would be more, resulting into thicker outer shell of the CD. The ratio $D_c/D_{CD}$ in Fig. 3 confirms this trend. The vortices would also trigger rotation of the anode-spot around the anode axis. The footmark of such anode-spot rotation is clearly visible in Fig. 2c.

During erosion, the anode is subjected to various heat and mass transfer mechanisms, where, the most dominant contributing factors are the convection, electron energy flow, radiation and sublimation [20]. The thermal conductivity has no practical stake in deciding any heat or mass exchange when the anode temperature approaches $T_s$.

The behaviour of the hydrodynamic boundary layer in between the anode and the plasma is governed by turbulent flow which leads to irregular, high frequency fluctuations of the flow properties. Theoretical analysis of this boundary layer is extremely difficult and cannot be handled without having prior knowledge of exact boundary conditions. However, this does not restrict one to have a universal consequence that is directly related to the nucleation and growth mechanism of CNTs and layered-graphene, as will be evident from the following discussion.

Upon reaching the temperature $T_s$, the carbon bonds of the anode starts softening, the effect being the maximum at the surface layers of each graphite grain in contact with the plasma. The non graphitic carbon contents forming the regions of grain boundaries and the graphitic carbon atoms adjacent to those boundaries have thus the maximum probability of getting dislodged from their sites and such atoms or atomic clusters leave the anode-surface as carbon vapour jet containing Cumulene structures, enriched mostly with $C_2$ (~0.04-0.12 mole fraction) and $C_3$ (~0.78-0.88 mole fraction) molecules if the carbon vapour temperature is close to 4000 ˚C, near atmospheric pressure [21]. These species originate respectively from the armchair and zigzag edges of the basal planes, in contact with the grain and quilt boundaries.

The vapour jet, thus generated, diffuses out of the anode both inside the plasma against the plasma pressure, as well as in the buffer medium. If the rate of production of carbon-vapour is larger than the rate of depletion of the same by diffusion (assuming no chemical reaction in the region) close to hydrodynamic boundary layer, the situation leads to formation of a dense carbon cloud in front of the anode. The resistive barrier, thus formed, not only slows down the drift velocity of the impinging electrons, but shields the heat flux from the plasma column to the anode too. As a result, the local Spitzer and thermal resistance increase to such an extent that the anode may practically be considered to be isolated from the arc channel and hence the portions of the anode, previously in contact with the electron channel stop eroding. At the same time, near the hydrodynamic boundary layer, electrons get populated and the anode voltage drop turns negative. So, in order to maintain a constant $I_{arc}$, the anode-spot, in case of CAM, very quickly move to random new non-eroding locations at regular intervals.

Due to this quick random movement of the anode-spot, different parts of the anode-surface undergo rapid thermal loading-unloading processes (which are as good as thermal shocks) leading to thermal fatigue to such extent, that the thermal shock resistance of graphite near the anode-surface gets sacrificed. While the regions, where anode-spot moves to, experience hot thermal shock (HTS), the rest of the regions are influenced either by HTS or cold thermal shock (CTS) depending on the local value of $\theta$, which is equal to $(T_a - T_p)$. $\theta$ is positive for CTS and negative for HTS.

In case of IAM, thermal shocks affect only the extreme peripheral region of the anode, with $\theta$ having much higher values than those in case of CAM. This is because; the temperature of the ambient is generally much lesser than that inside the plasma column.

Though being absolutely random, $\theta$ varies with $I_{arc}$, crystallographic nature of the eroding surface, $P_o$, cooling scheme, arc-length and the heat transfer coefficients of the buffer gas and plasma medium, along with their flow patterns. Thus, estimation of $\theta$ requires extensive computation, which may not be possible to execute right now, because of insufficient data and lack of appropriate model(s) that can handle such a complex situation. Still, it is obvious that, in case of CAM, the higher the value of $I_{arc}$, lesser would the variation in $\theta$.

The effects of the thermal shocks would be as follows. The pre-existing cracks (preferably nearly parallel to the eroding surface) may extend through the bulk of the graphite near the surface under the influence of HTS, leading to fragmentation of the crystallites into smaller pieces, which are detached from the rest of the anode body and move downstream. The remnants of these crystallites are always found within the CDs [21].

On the other hand, if a CTS takes place near a surface groove, an intergranular steady crack ($C_{r0}$) may originate at the anode-surface and penetrate into the bulk with microbranches ($C_{r1}s$). Such a failure of polycrystalline graphite has been discussed by the author elsewhere [22]. The possibility of such crack propagation depends on whether the relative orientations of the adjacent grains support the bifurcation of the propagating intergranular crack at regular intervals or not [22]. Thus, the grain boundary character distribution (GBCD) within the anode is of extreme importance, while analyzing the post CTS phenomena.

If the above mentioned microbranching instability takes place, $C_{r1}$s cleave the adjacent crystallites parallel to the basal planes (Fig. 5) into rotationally misoriented graphitic nano ribbons (GNRs) [22], with an average thickness $x$, say. Here, $C_{r0}$ can be viewed as a mode-I edge crack expanding under uniaxial plane-stress from the surface to interior of the anode. Under such a condition, the corresponding stress intensity factor $K_I$ reduces to $\sigma_T\sqrt{\pi l_c}$ [23], where $\sigma_T$ and $l_c$ are the thermal stress acting on the fracturing zone and the semi-crack-length of $C_{r0}$ respectively. Microbranching of a $C_{r0}$ at a critical stress intensity factor $K_{IC}$ and crack-length $l_c$ may be initiated inside the feedstock graphite when $C_{r0}$, after covering a minimum distance $l_c$, releases its energy at a rate $G_c$ per unit area to form four new crack surfaces parallel to the (002) planes on both the sides of that $C_{r0}$. Using the concepts of elastic continuum, Griffith's theory of brittle fracture [24], and the analysis by Irwin [25], one can write:

$$G_c = 4\gamma' = \frac{K_{IC}^2}{E'} \qquad (2),$$

where $\gamma'$ and $E'$ are the energy required to tear apart two adjacent basal planes per unit surface area created by $C_{r1}$ and elastic modulus of graphite crystallites along $c$ axis respectively. Using Eqn. 2, one thus gets the following limiting condition, in which $l_c$ is nothing but $x$ corresponding to a particular set of operating parameters.

$$K_{IC} = \sigma_T\sqrt{\pi x} = 2\sqrt{E'\gamma'} \qquad (3)$$

Plugging in the expression of the thermal stress $\sigma_T$ and approximate values of the thermo-elastic parameters from the literature [22], in Eqn. 3 and simplifying one obtains

$$x = (N-1)d_{002} = \frac{4\gamma'}{\pi\alpha'^2\theta^2E'} = \sim\frac{10^6}{\theta^2}\,\text{Å} \qquad (4),$$

with $\theta$ measured in °C or K and $\alpha'$, $N$ and $d_{002}$ being the coefficient of thermal expansion of anode grains along $c$ axis, the average number of basal planes present in between two consecutive $C_{r1s}$ and the inter-planer separation of the basal planes of graphite (~3.35 Å at room temperature) respectively.

The generation of two surfaces by $C_{r1}$ is possible only if the thickness of $C_{r1}$ is ~7 Å, a distance at which the van der Waals interaction in between the associated basal planes becomes negligible [26]. It immediately follows from this understanding that the minimum distance in between the tips of two consecutive parallel $C_{r1}$s cannot be lesser than ~7 Å. Thus, the minimum number of layers, peeled off by $C_{r1}$s, cannot be lesser than 2 (refer to Fig. 5).

While propagating, $C_{r1}s$ remain sub-critical, the local environment of $C_{r1}$ remains linearly elastic and thus, the crack tip is not atomistically sharp; rather it is blunt with a finite radius of curvature $\rho$. The propagation of $C_{r1}$ being a spontaneous process, there is no control on the value of $\rho$.

At the tip of a $C_{r1}$ two stress field components (Fig. 5) are active; while, $\vec{\sigma_f}$ drives the crack in the forward direction, $\vec{\sigma_t}$ tears apart the adjacent basal planes. As a result, the resultant stress $\vec{\sigma_c}$ bends the set of basal planes, the outer radius of curvature ($R_o$) of which depends on the flexural rigidity ($D$) of the bent sheets, $\rho$ and $d_{002}$. It is thus clear that $R_o \propto \frac{Dd_{002}(T_a)}{\rho}$. The bending stress, near the tips of $C_{r1}$s breaks the GNRs into tiny curved fragments along the lines of lattice discontinuity. These curved GNRs (c-GNRs) later serve as the seeds for the nucleation of both CNTs and graphitic flakes including few-layer graphene sheets (FLGs).

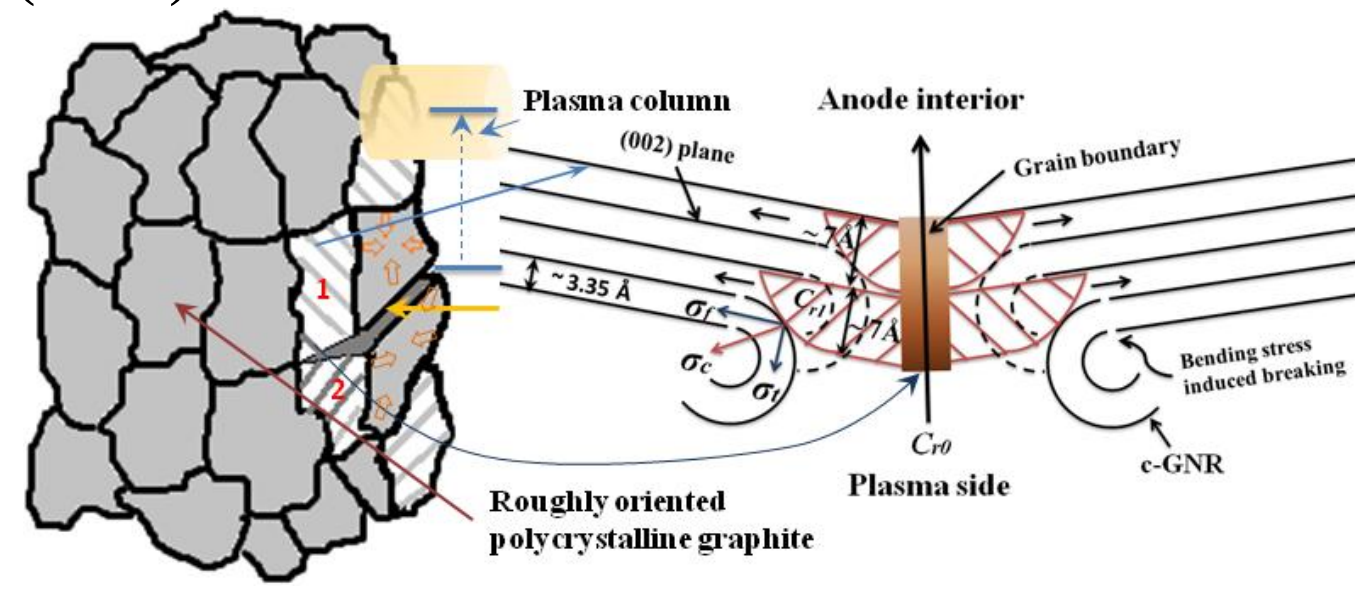

**Fig. 5** Conceptual schematic depicting brittle fracture in polycrystalline graphite. The figure is a 2D representation of the microbranching of a Griffith's crack propagating along a grain boundary, in a polycrystalline graphite anode, on the left of the figure. A shift in the plasma column, from the location of the grains 1 and 2, as indicated by the dotted arrow, induces rapid contraction in the grains. A crack originates at the surface grain boundary (indicated by the orange arrow) of the grains 1 and 2, the microbranches of which cleave and fragment the grains into c-GNRs, as indicated on the right hand side of the figure. The schematic corresponds to the minimum possible separation in between two consecutive microbranches.

The c-GNRs, after being released to the adjacent gaseous medium, undergo three competitive processes, which would decide their final forms. (i) They try to spring back to reduce their stored elastic strain energy, (ii) the viscosity of the vapour ambience tries to oppose the springback and (iii) the $C_2$ and $C_3$ species present in the vapour try to chemically react with their open armchair and zigzag edges respectively to reduce the Gibb's free energy of the system.

In case of CAM, c-GNRs are released in the inter-electrode space having high vapour pressure and temperature. Hence, the viscosity of the medium is high and the c-GNRs cannot springback before transforming into seamless open-ended cylinders with outer radius $R_o$ (hereafter termed as p-CNTs) by chemically reacting with the $C_2$ and $C_3$ species (as explained in Fig. 6). Such a transformation is energetically favourable as the Tersoff's potential of the system attains minima under such a condition [27]. The p-CNT, thus generated, can further grow only along its axial direction [28], as schematically illustrated in Fig. 6.

Because of such formation of p-CNTs, it is also clear that if one compares the inner radii ($R_i$) of two as synthesized CNTs with almost equal value of $R_o$, smaller is the value of $R_i$ for that CNT, which has more number of walls. This conclusion is in excellent agreement with the very first TEM images of arc-generated CNTs, available in the literature [1].

It is clear from the above discussion that under CAM, the probability of finding the zones, conducive to microbranching failure, increases with $I_{arc}$ in case of roughly oriented polycrystalline graphite. Such enhanced probability would likely to increase the tube to particle ratio in the system. As this ratio is also proportional to the $\frac{I_G}{I_D}$ ratio [15], one should find a steady rise in $\frac{I_G}{I_D}$ ratio with $I_{arc}$. Fig. 3 supports this fact very well.

In contrary to the above situation, during IAM operation, the c-GNRs are formed at the side-surface of the anode, which remains in direct contact with the much colder ambience with much lower pressure than that of the carbon vapour present inside the discharge region. These c-GNRs, after getting detached from the anode, are thus exposed to a low viscous environment having both carbon and buffer gas molecules. As a result, majority of the as-generated c-GNRs cannot transform into p-CNTs; rather they springback and grow as flakes outside the discharge zone and get deposited on the reactor chamber walls [29]. Some of them are dragged inside the discharge region by the intense-arc generated vortex. Inside the discharge zone, the GNRs lose their edge-atoms by vaporization while passing through the region having $T_p > T_s$. Close to the cathode, they again grow along their edges by reacting with the $C_2$ and $C_3$ species before getting deposited inside the core of the CD.

During IAM operation, the value of $\theta$ is high. As a result, following Eqn. 4, the flakes formed during IAM, would contain only a couple of basal planes. Most of these flakes would therefore appear mostly as few or multi layer graphene [3].

The assumption, that both the CNTs and layered-graphene would grow at temperatures lower than $T_s$, is valid on account of well accepted experimental evidences [30].

During the growth of GNRs, the reactivity of its constituent graphene layers are not same due to curvature induced strain [31]. As a result, the basal plane with the smallest radius of curvature, grows more than other adjacent planes. So, in the as-synthesized flakes, there might be edge-regions, which would appear like a monolayer graphene. Probably, this is why Wu *et al*. observed monolayer graphene in a similar arcing process [32].

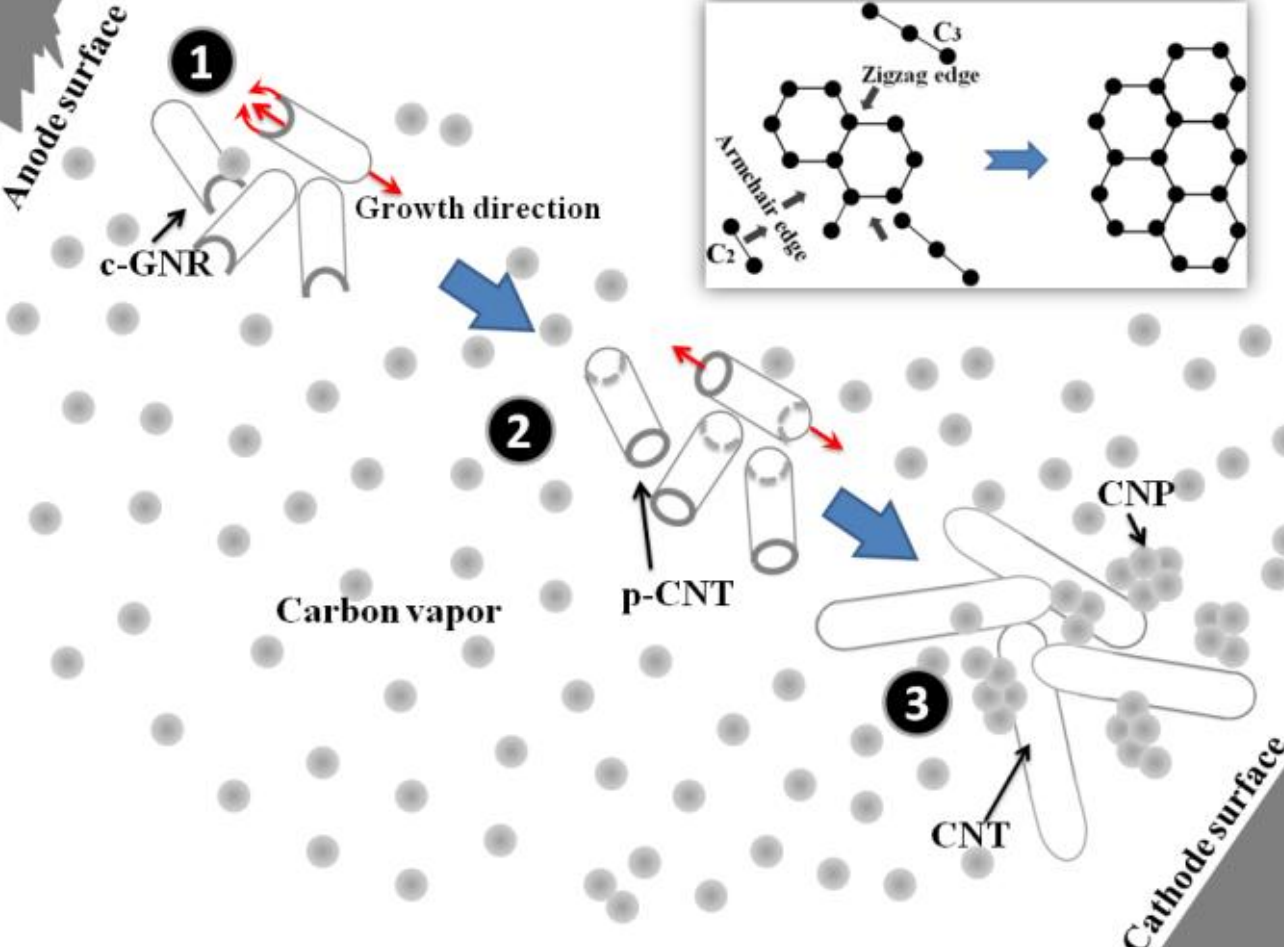

**Fig. 6** Conceptual schematic representing the nucleation and growth of CNTs inside carbon arc plasma. At step 1, c-GNRs are formed from the anode and released to the carbon vapour ambience. At step 2, these c-GNRs grow along the open edges by chemically reacting with $C_2$ and $C_3$ molecules present in the plasma and transform into p-CNTs, as they migrate towards the colder cathode. The red arrows indicate the directions of growth. The inset shows how individual basal planes in the c-GNRs grow in dimension. At step 3, p-CNTs are fully grown and their growth is terminated by the capping of their open ends. Near the cathode, Cumulene structures, present in the plasma, mutually react with each other to finally get condensed in the forms of CNPs or amorphous slurry.

During the growth of p-CNTs, they try to orient themselves along the arc electric-field-lines due to induced electric dipole moments and migrate towards the cathode. As a result, these CNTs are not found on the reactor chamber walls.

As a dense carbon vapour is essential for the nucleation and growth of CNTs, the arc-plasma can therefore be treated as a solid/plasma colloid, where growing p-CNTs undergo Brownian motion. The corresponding motion and hence the time of flight of a growing p-CNT are similar to random walk amidst a large number of particles leading inevitably to collisions: the Brownian coagulation. Thus, the effective distance travelled by a p-CNT during the Brownian motion decides the final length of the CNT, provided the growth is not terminated midway on account of capping of both of its ends, following the mechanisms discussed elsewhere [33]. This two-phase colloid being thermodynamically unstable with respect to coarsening process, flocculation is provoked in such a system resulting in bundles of CNTs near the cathode region. The fact that arc-generated CNTs always appear in bundles, justifies the validity of the proposed mechanism.

Arc-generated CNTs are always accompanied by a large number of particles with variable shapes and sizes. There are mainly two types of contributors in this regard. While the HTS generated remnants form the first group, the second one is formed by the particles nucleated within the arc as a result of gas-phase condensation of the Cumulene structures.

The particles, nucleated from the Cumulene molecules, often lack a well-defined crystalline structure and some of them may finally condense on the CNTs or the electrode-surfaces in the forms of slurry, globules or even rippled thin films.

## Conclusions

In a graphite arc-discharge, the nucleation of both CNTs and layered-graphene is initiated by GNRs eroded from the polycrystalline graphite anode on account of CTS. The nucleation and growth of CNTs are promoted, when the arc is constricted and the electrodes are placed very close to each other. In case of an intense arc, the system would produce mostly flaky structures, a large fraction of which might be FLGs. As the minimum number of basal planes in a c-GNR cannot be 1 (following Eqn. 4) and has to be 2 (Fig. 5), it is never possible to synthesize either single walled CNT or mono-layer graphene in a pure graphite electric-arc process. Both the arc-generated CNTs and layered-graphene would be turbostatic, as CTS induced GNRs are found to be composed of rotationally misoriented graphene sheets [22]. The yield and morphology of these CNTs greatly depend on the internal structure of the anode and associated GBCD. If the basal planes are parallel to the eroding surface, CNTs would nucleate and grow to the maximum. However, if these planes are orthogonal to the eroding surface, the formation of CNTs would be minimal. These predictions are in excellent agreement with the experimental observations available in the literature [12]. The number of walls in the as-grown CNTs and the number of layers in the flakes would inversely vary with $\theta^2$, which would have a range of variation on the anode surface, even if all operating conditions are kept constant. The length of the CNTs and the dimension of the layered-graphene can be controlled by regulating the time of flight of the p-CNTs and GNRs within the inter-electrode gap. The diameter distribution of these CNTs may be manipulated by controlling the ambient gas cooling, which would decide $T_a$ close to the c-GNR forming zones at the onset of microbranching instability. Successful experimentation in this direction [34], substantiates the validity of the proposed model. The loss of feedstock in the form of graphite chunks is related to the presence of the pre-existing internal cracks that grow as a result of HTS. This unwanted loss, thus may be minimized by eroding a high density graphite sample.

In a nutshell, the article predicts that, in order to maximize the conversion efficiency of the feedstock graphite into CNTs or FLGs, in an arc-discharge process, the most important challenge ahead is to engineer the internal structure of eroding electrode. For tailoring the structures of these nanomaterials, one needs to have mastery on controlling both the fracture dynamics inside the eroding electrode and the composition of the plasma column.

## Conflicts of interest

There are no conflicts to declare.

## Acknowledgements

It is really gratifying that Prof. (Mrs.) S. V. Bhoraskar kindly read the manuscript and gave useful suggestions to improve the quality of the paper. The author thanks Dr. Vikas L. Mathe and the Department of Physics, Savitribai Phule Pune University for allowing the author to generate the data presented in the paper. Thanks are due to Dr. V. G. Sathe, UGC-DAE Center for Scientific Research, Indore for recording the Raman spectroscopic data. Special thanks are due to Late Dr. Naveen V. Kulkarni for his kind assistance during the operations of the arc reactor.

## References

[1] S. Iijima, Helical microtubules of graphitic carbon. *Nature* **354**, 56-58 (1991).

[2] V. N. Popov, Carbon nanotubes: properties and application. *Mat. Sci. Eng. R* **43**, 61–102 (2004).

[3] S. Karmakar, A. B. Nawale, N. P. Lalla, V. G. Sathe, S. K. Kolekar, V. L. Mathe, A. K. Das, S. V. Bhoraskar, Gas phase condensation of few-layer graphene with rotational stacking faults in an electric-arc. *Carbon* **55**, 209–220 (2013).

[4] A. K. Geim, Graphene: status and prospects. *Science* **324**, 1530–1534 (2009).

[5] E. G. Gamaly, T. W. Ebbesen, Mechanism of carbon nanotube formation in the arc discharge. *Phys. Rev. B* **52**, 2083—2089 (1995).

[6] Y. W. Liu, L. Wang, H. Zhang, A possible mechanism of uncatalyzed growth of carbon nanotubes. *Chem. Phys. Lett.* **427**, 142–146 (2006).

[7] S. Iijima, Growth of carbon nanotubes. *Mat. Sci. Engg. B* **19**, 172–180 (1993).

[8] W. A. de Heer, P. Poncharal, C. Berger, J. Gezo, Z. Song, J. Bettini, D. Ugarte, Liquid Carbon, Carbon-Glass Beads, and the Crystallization of Carbon Nanotubes. *Science* **307**, 907—910 (2005).

[9] V. Gupta, Graphene as intermediate phase in fullerene and carbon nanotube growth: A Young–Laplace surface-tension model. *Appl. Phys. Lett.* **97**, 181910-1— 181910-3 (2010).

[10] P. J. F. Harris, Solid state growth mechanisms for carbon nanotubes. *Carbon* **5**, 229–239 (2007).

[11] R. E. Smalley, From dopyballs to nanowires. *Mater. Sci. Eng. B* **19**, 1—7 (1993).

[12] J. M. Lauerhaas, J. Y. Dai, A. A. Setlur, R. P. H. Chang, The effect of arc parameters on the growth of carbon nanotubes. *J. Mat. Res.* **12**, 1536–1544 (1997).

[13] M. Togaya, Pressure dependences of the melting temperature of graphite and the electrical resistivity of liquid carbon. *Phys. Rev. Lett.* **79**, 2474– 2477 (1997).

[14] M. Musella, C. Ronchi, M. Brykin, M. Sheindlin, The molten state of graphite: an experimental study. *J. Appl. Phys.* **84**, 2530-2537 (1998).

[15] S. Karmakar, N. V. Kulkarni, V. G. Sathe, A. K. Srivastava, M. D. Shinde, S. V. Bhoraskar, A. K. Das, A new approach towards improving the quality and yield of arc-generated carbon nanotubes. *J. Phys. D: Appl. Phys.* **40**, 4829–4835 (2007).

[16] S. Karmakar, N. V. Kulkarni, A. B. Nawale, N. P. Lalla, R. Mishra, V. G. Sathe, S. V. Bhoraskar, A. K. Das, A novel approach towards selective bulk synthesis of few-layer graphenes in an electric arc. *J. Phys. D: Appl. Phys.* **42**, 115201-1—14 (2009).

[17] P. Y. Huang, C. S. Ruiz-Vargas, A. M. van der Zande, W. S. Whitney, S. Garg, J. S. Alden, C. J. Hustedt, Y. Zhu, J. Park, P. L. McEuen, D. A. Muller, Grains and grain boundaries in single-layer graphene atomic patchwork quilts**.** *Nature* **469**, 389–392 (2011).

[18] N. S. Rasor, J. D. McClelland, Thermal properties of graphite, molybdenum and tantalum to their destruction temperatures. *J. Phys. Chem. Solids* **15**, 17–26 (1960).

[19] S. Gershman, Y. Raitses, Unstable behavior of anodic arc discharge for synthesis of nanomaterials. *J. Phys. D: Appl. Phys.* **49**, 345201-1–9 (2016).

[20] J. McKelliget, J. Szekely, Heat transfer and fluid flow in the welding arc. *Metall. Trans. A* **17**, 1139-1148 (1986).

[21] J. Abrahamson, Graphite sublimation temperatures, carbon arcs and crystallite erosion. *Carbon* **12**, 111—141 (1974).

[22] S. Karmakar, A. B. Nawale, N. S. Kanhe, V. G. Sathe, V. L. Mathe, S. V. Bhoraskar, Tailored conversion of synthetic graphite into rotationally misoriented few-layer graphene by cold thermal shock driven controlled failure. *Carbon* **67**, 534—545 (2014).

[23] D. P. Rooke, D. J. Cartwright, *Compendium of stress intensity factors. Ministry of Defence. Procurement Executive. London: H. M. S. O.* (1976).

[24] A. A. Griffith, The phenomena of rapture and flow in solids. *Phil. Trans. R. Soc. Lond. A* 221, 163—198 (1921).

[25] G. R. Irwin, Crack-extension force for a part-through crack in a plate. *J. Appl. Mech.* **29**, 651–654 (1962).

[26] L. A. Girifalco, M. Hodak, Van der Waals binding energies in graphitic structures. *Phys. Rev. B* **65**, 125404-1–125404-5 (2002).

[27] S. Sawada, N. Hamada, Energetics of carbon nano-tubes. *Solid State Comm.* **83**, 917–919 (1992).

[28] A. Maiti, C. J. Brabec, C. Roland, J. Bernholc, Theory of carbon nanotube growth. *Phys. Rev. B* **52**, 14850—14858 (1995).

[29] Y. Wu, B. Wang, Y. Ma, Y. Huang, N. Li, F. Zhang, Y. Chen, Efficient and large-scale synthesis of few-layered graphene using an arc-discharge method and conductivity studies of the resulting films. *Nano Res.* **3**, 661-669 (2010).

[30] H. M. Yusoff, R. Shastry, T. Querrioux, J. Abrahamson, Nanotube deposition in a continuous arc reactor for varying arc gap and substrate temperature. *Curr. Appl. Phys.* **6**, 422–426 (2006).

[31] T. Dumitrică, C. M. Landis, B. I. Yakobson, Curvature-induced polarization in carbon nanoshells. *Chem. Phys. Lett.* **360**, 182-188 (2002).

[32] C. Wu, G. Dong, L. Guan, Production of graphene sheets by a simple helium arc-discharge. *Physica E* **42**, 1267–1271 (2010).

[33] W. Z. Zhu, D. E. Miser, W. G. Chan, M. R. Hajaligol, Characterization of multiwalled carbon nanotubes prepared by carbon arc cathode deposit. *Mat. Chem. Phys.* **82**, 638–647 (2003).

[34] S. Akita, S. Kamo, Y. Nakayama, Diameter control of arc produced multiwall carbon nanotubes by ambient gas cooling. *Jpn. J. Appl. Phys.* **41**, L 487–L 489 (2002).